\documentclass{article}
\usepackage[english]{babel}
\usepackage{amsfonts,amssymb, amsmath}

\newtheorem{prop}{Proposition}

\begin{document}

\title{On the Darboux-Halphen system: Jacobi vs Lie}
\author{A.\,V.~Tsiganov\\
\it\small St. Petersburg State University, St. Petersburg, Russia;\\
\it\small Beijing Institute of Mathematical Sciences and Applications, Beijing, China;\\
\it\small e--mail: andrey.tsiganov@gmail.com
}
\date{}
\maketitle

\begin{abstract}
Two constructions of the Darboux-Halphen system are discussed. In the Jacobi construction we start with transcendental  functions which are fixed as the first integrals. In the Lie construction we use a single-valued representation of the simple Lie algebra $sl(2,\mathbb R)$ which is non-integrable in Lie's terminology.  
\end{abstract}

\section{Introduction}
\setcounter{equation}{0}
The Darboux-Halphen system 
\begin{equation}\label{halp-sys}
\left\{ \begin{array}{l}
\dot{x}_1=x_2x_3-x_1(x_2 + x_3)\,,\\
\dot{x}_2=x_1x_3-x_2(x_1+x_3)\,,\\
\dot{x}_3=x_1x_2-x_3(x_1+x_2)\,.
\end{array}
\right.
\end{equation}
first appeared in Darboux’s work on triply orthogonal surfaces \cite{darb-3}. The general solution to (\ref{halp-sys}) was determined in 1881 by  Halphen \cite{hal,hal2} and Brioschi \cite{bri}  in terms of the elliptic modular function; that is, the square of the elliptic modulus, viewed as a function of the ratio $\tau$ of the periods of the Jacobi elliptic functions.  The physical contexts in which these equations have appeared include dynamics of pairs of magnetic monopoles, theory of vacuum Einstein equations for hyperkähler Bianchi IX metrics,  similarity reductions of associativity equations on a three-dimensional Frobenius manifold \cite{chak99}.

A system of differential equations is said to be 'integrable' if there is a single-valued solution in the neighbourhood of any given point that satisfies the specified conditions. These solutions may be complete, in quadratures, meromorphic, or algebraic.  The classical Darboux-Halphen system and its known generalizations are integrable in the sense that their complete solutions can be expressed in terms of two linearly independent solutions of certain hypergeometric equations. However, some generalizations do not possess the Painlev\'{e} property in the usual sense because their solutions admit a movable natural barrier instead of isolated singular points, see \cite{chak99,chak03,chak18,chak25} and references within.

All classical integrability theorems are concerned with complete integrability: Euler's integrability condition \cite{eul-int}, Pfaff theorem \cite{pfaff}, Euler-Jacobi theorem \cite{jac}, Clebsch-Deahna-Frobenius theorem \cite{cl61,cl66,dea,frob}, Donkin-Liouville theorem about commutative integrability \cite{don,liu}, Lie's theorem on noncommutative integrability \cite{lie-book}, Darboux theorem \cite{darb-pfaff}, etc. The corresponding complete solutions are local solutions that depend on the necessary number of arbitrary constants. By resolving these complete solutions with respect to these constants, we obtain local first integrals for the given complete integrable system. A more rigorous treatment can be found in the Carath\'{e}odory textbook \cite{carat}, pages 23-25.

So, all theorems concerning complete integrability are local, and the corresponding first integrals may be local, multivalued, or transcendental functions according to Euler's terminology. Since, there are many constructions of complete integrable systems without exact calculation of the integral functions  \cite{lie-book, for90, lie-book2, car,kah}. Various such constructions were proposed by Grassmann, Hamburger, Deanha, Clebsch, Natani, Mayer, Lie, Frobenius, Darboux, Cartan, etc. In the context of modern topological mechanics, multi-valued functionals and non-exact closed differential 1-forms are discussed in \cite{nov82}.

Our aim is to compute Darboux-Halphen-type polynomial differential systems using the classical Jacobi and Lie constructions of complete differential systems.  Both constructions are applicable to complete differential systems with transcendental or multi-valued first integrals. The main difference is that in the Jacobi method, we start with the multi-valued functionals, whereas in the Lie method, we start with a single-valued realisation of a real Lie algebra.

\subsection{Jacobi's complete differential systems}
In an important and influential paper of 1827 \cite{jac-27} Jacobi studied a homogeneous linear first-order partial differential equation
\begin{equation}\label{pdejac-eq}
X_1(x_1,\ldots,x_n)\frac{\partial z}{\partial x_1}+\cdots+X_n(x_1,\ldots,x_n)\frac{\partial z}{\partial x_n}=0\,.
\end{equation}
The integration of this equation was considered to be equivalent to integrating the system of ordinary differential equations
\begin{equation}\label{mjac-eq}
{\mathrm d}x_1:\mathrm{d}x_2:\cdots:\mathrm{d}x_n=X_1:X_2:\cdots:X_n\,.
\end{equation}
This form of equations Jacobi introduced to indicate that any one of the variables $x_i$ could be chosen as independent variable.  If $x_n$ is chosen we have $n-1$ equations
\begin{equation}\label{njac-eq}
\frac{\mathrm{d}x_i}{\mathrm{d}x_n}=\frac{X_i}{X_n}\,,\qquad i=1,\ldots,n-1\,.
\end{equation}
If an auxiliary variable $t$ is introduced into (\ref{mjac-eq}) by setting
the common ratio there equal to $\mathrm d t$, then (\ref{mjac-eq}) may also be written in the form
\[
\frac{ \mathrm{d}x_i}{\mathrm{d}t}= X_i(x_1,\ldots, x_n), \qquad i = 1,\ldots, n.
\]
To  prove equivalence of the homogeneous partial differential
equation (\ref{pdejac-eq}) and the system of ordinary differential equations (\ref{mjac-eq}-\ref{njac-eq}), Jacobi
considered complete solutions of (\ref{njac-eq}) 
\begin{equation}\label{intjac-eq}
x_k = \phi_k(x_n,C_1,\ldots,C_{n-1}),\qquad  k = 1, \ldots, n-1\,,
\end{equation}
corresponding to arbitrary initial conditions $x_k = C_k$ when $x_n=x_{n}^{0}$, see discussion in  \cite{lie-book}. 

By classical Jacobi's definition equations (\ref{njac-eq}) are completely integrable when  functionally independent $n-1$ functions  $\phi_k (x_n,C_1,\ldots,C_{n-1})$  are known. If  ordinary differential equations (\ref{njac-eq}) are completely integrable,  we can solve  equations (\ref{intjac-eq}) for the $C_k$ and obtain
\begin{equation}\label{fk-jac}
 C_k = f_k(x_1,\ldots,x_n)\,,\qquad k=1,\ldots,n-1\,.
\end{equation}
If the equations $f_k = C_k$ can be  differentiated with respect to $x_n$, each function $z =f_k(x_1,\ldots,x_n)$ is then a solution to partial differential equation (\ref{pdejac-eq}) which is also called complete equation in this case. 

Thus, for Jacobi and his successors the equations (\ref{pdejac-eq},\ref{mjac-eq}) and (\ref{njac-eq}), were regarded as equivalent, and they would switch from one to the other whenever convenient. In the generic case  solutions $f_k(x)$ of equations (\ref{fk-jac}) are multi-valued functions on the variables $x_k$ even though the complete solutions $\phi_k$ (\ref{intjac-eq}) are single-valued.

\subsection{Jacobi's construction of complete differential systems}
Using Proposition at the end of the page 331 of tome 4 of Jacobi's Œuvres compl\`{e}tes \cite{jac}, we can directly compute a complete system of ordinary differential equations (\ref{mjac-eq})
\begin{equation}\label{ode-1}
\frac{\mathrm{d}x_1}{X_1}=\frac{\mathrm{d}x_2}{X_2}=\cdots=\frac{\mathrm{d}x_n}{X_n}
\end{equation} 
using arbitrary independent functions  $f_1, f_2,\ldots, f_{n-1}$ on $n$ variables $x_1,\ldots,x_n$ and multiplier $M(x)$. Here  $X_k(x_1,\ldots,x_n)$ are defined by equation (1) on the page 331   
\begin{equation}\label{jac-1}
MX_k=A_k\,,\qquad k=1,\ldots,n\,,
\end{equation}
where $M\neq 0$ and $A_k$ are minors of the functional determinant
\[
R =\sum \pm \frac{\partial f}{\partial x_1}\frac{\partial f_1}{\partial x_2}\cdots\frac{\partial f_{n-1}}{\partial x_n}
= A_1\frac{\partial f}{\partial x_1}+A_2\frac{\partial f}{\partial x_2}+\cdots+
A_n\frac{\partial f}{\partial x_n}\,,
\]
see equation (2) and last equation on the page 331. In the modern notation of Jacobi's functional determinants  we have to put $f=x_k$ for varying $k$, i.e. 
\[
A_k=\frac{\partial(x_k,f_1,\ldots,f_{n-1})}{\partial(x_1,x_2,\ldots,x_n)}\,.
\]
see the Goursat textbook \cite{gour91}.

By definition (\ref{jac-1}) multiplier $M$ satisfies the equation (3) on the page 331
\begin{equation}\label{jac-3}
\frac{\partial MX_1}{\partial x_1}+\cdots+ \frac{\partial MX_n}{\partial x_n}=0\,,
\end{equation}
according to the property of functional determinants found by Jacobi \cite{jac-book}. 

The Jacobi definition of a complete solution, along with his two methods of constructing equations with complete solutions, formed the basis for subsequent research by Donkin, Liouville, Clebsch, Lie, Frobenius, Cartan and Darboux, see classical textbooks \cite{for90,lie-book2,car,kah,im69,gour91}.  

As example, in \S24 of the Lie textbook \cite{lie-book} we can find definition of complete solution, complete equations and new construction of complete differential equations based on the groups of infinitesimal transformations. In Cartan's textbook \cite{car} Jacobi's construction of vector field $X$ (\ref{jac-1}) was rewritten using a set of known invariant differential forms $\omega_i$ and invariants completely antisymmetric unit tensor field 
\[\mathcal E= \partial_1\wedge\partial_2\wedge\cdots\wedge\partial_n\,,\qquad\mbox{where}\qquad \partial_i=\frac{\partial}{\partial x_i}\,,\]
so that 
\[
MX=\mathcal E \omega_1\cdots \omega_{n-1}\,.
\]
It allows us to rewrite vector field $X$ in the formal Hamiltonian form $X=P_i\omega_i$ using the
rank-two bivectors 
\[
P_i=(-1)^{n-i-1}M^{-1} \mathcal E\, \omega_1\dots  \widehat{\omega}_i \dots \omega_{n-1}\,,\qquad i=1,\ldots,n-1\,,
\]
the symbol $\widehat{\omega}_i$ means that the one-form $\omega_i$ is missing from the product of one-forms. 

For $n=2$,  Jacobi's construction coincides with the Euler construction of integrable planar differential systems
\begin{equation}\label{eq-pf}
\omega=P(x, y)dx+Q(x, y)dy=0\,,
\end{equation}
where $P(x, y)$ and $Q(x, y)$ are given by
\[
MP=\frac{\partial \varphi}{\partial x} \qquad\mbox{and}\qquad MQ=\frac{\partial \varphi}{\partial y}\,.
\]
In Euler's textbook \cite{eul-int} we can find many examples when $\varphi(x,y)$ and $M(x,y)$ are multi-valued functions which give rise to single-valued differential equations (\ref{eq-pf}). 
As an example, in \S471-\S472 Euler studied first integral
\[
\varphi(x,y)=\frac{y^{1-n}}{1-n}\,e^{(1-n)\int Xdx}-\int e^{(1-n)\int Xdx} \mathfrak{X}\,dx\,.
\]
which gives rise to differential equation
\[
dy+Xydx=\mathfrak{X}y^ndx\,,
\]
where $X$ and $\mathfrak{X}$ are single-valued functions on $x$. In the generic case both the integral $\varphi$ and the multiplier \[M=  \int e^{(1-n)\int Xdx}\] 
are multi-valued or transcendental functions in Euler's terms. 

The Jacobi theory involves functions $f_k(x_1,\ldots, x_n)$ of any number of variables, the properties of which are not specified explicitly. It is not even clear whether the variables are assumed to be real or complex.
However the equality of mixed partial derivatives, the inverse function theorem and the implicit function theorem are applied whenever needed. During this period, mathematicians generally regarded variables as complex rather than real, as this assumption allows for the existence of partial derivatives of these functions and facilitates the frequent use of the equality of mixed partial derivatives.
Whether Jacobi was fully aware of the local nature of his results is far less certain, but readers should understand them as local results that are valid in the neighbourhood of any point that satisfies the specified conditions. It was not until the 20-th century that mathematicians began to take the distinction between local and global results seriously.

\subsection{Lie's construction of complete equations}
The intense and widespread interest in the theory of first-order partial differential equations, which was triggered by the publication of Jacobi's theory, is reflected in the publication of Imschenetsky's monographic essay \cite{im69}. According to Klein's recollections, it seems to have been the primary source from which Lie learned this theory, as Lie studied Imschenetsky's essay with great enthusiasm while he and Klein were together in Paris in 1870.

In \cite{lie-book,lie-book2} Lie proposed to  construct invariant complete differential systems  (\ref{ode-1}) solving equations (\ref{lie-cl}) for the given  infinitesimal transformations  group.
The central concept in Lie's vision of a geometrically informed approach to the study of partial differential equations was that of an  infinitesimal transformation group. Following Lie we suppose that system of equations (\ref{ode-1}) admits a set of infinitesimal transformations 
\[
Y_i(f) =\sum \xi_{i,j}(x_1,\ldots,x_n)\partial_j (f)\,.
\]
If infinitesimal transformations $Y_i(f)$ form a group
\begin{equation}\label{lie-cl}
Y_i(Y_j(f))-Y_j(Y_i(f))=\sum c_{ij}^k Y_k(f)
\end{equation}
then the corresponding system of Pfaffian equations has a complete solution, see \cite{lie-book} and references on the Clebsch and Mayer papers within. Here, the $c_{ij}^k$ satisfy the Jacobi identities and may be either constants or functions of the first integrals, see \cite{lie-book,lie-book2} and Cartan's textbook  \cite{car}.

In Cartan's textbook the equivalent proposition about complete integrability of Pfaffian equations was formulated in terms of differential 1-forms and, therefore, called the Frobenius theorem. It is historically somewhat misleading to attach only Frobenius’s name to  this theorem, since the above theorem is just the dual of the Jacobi-Clebsch theorem from \cite{cl61,cl66}, see careful description of known results and references in \cite{frob} and textbook \cite{for90}. Furthermore, although Frobenius also gave a proof of the theorem that is independent of the Jacobi-Clebsch theorem and the consideration of partial differential equations, that proof was, as he explained on page 291 in \cite{frob}, simply a more algebraic and “symmetrical” version of one by Deahna \cite{dea}. Sadly, both Frobenius and Cartan fail to mention Mayer and Lie results, for instance  the new Lie definition of complete solution, see \S24 in \cite{lie-book}. On the other hand, in \cite{lie-book,lie-book2} Lie refers to Jacobi, Clebsch and Mayer, but completely ignores Frobenius results. 

In \cite{lie-75} and  \S 67 of textbook \cite{lie-book} Lie considers simply transitive groups  which contain $n$ independent infinitesimal transformations $Y_1,\ldots,Y_n$ for which  determinant
\begin{equation}\label{m-lie}
\Delta=\left\| 
\begin{array}{ccc}
\xi_{1,1}(x)&\quad\cdots \quad  &\xi_{1,n}(x)\\
\vdots&\vdots &\vdots\\
\xi_{n,1}(x)&\quad\cdots \quad &\xi_{n,n}(x)\\
\end{array}
\right\|\,,
\end{equation}
does not vanish identically. In this case function
\begin{equation}\label{mul-lie}
M=\Delta^{-1}
\end{equation}
satisfies the Jacobi equation (\ref{jac-3})
\[
\mbox{div} (MY_i)=0\,,\qquad i=1,\ldots,n,
\]
for all the infinitesimal transformations $Y_1,\ldots,Y_n$. So, we have $n$ independent complete partial differential equations
\[
Y_i(f)=0
\]
with the common multiplier $M=\Delta^{-1}$.  If we have $r<n$ independent infinitesimal transformations $Y_i$ and (\ref{lie-cl}) holds, the corresponding $r$ complete partial differential equations $Y_i(f)=0$ have different multipliers \cite{lie-book}.

If infinitesimal transformation group is integrable we can obtain the corresponding complete solution by using rectifying coordinates \cite{lie-book,lie-book2}. In modern literature authors use term "solvable Lie algebras" instead Lie's term "integrable infinitesimal transformation group". 

Let us consider integrable planar differential system (\ref{eq-pf})
\[
\omega=P(x, y)dx+Q(x, y)dy=0\qquad \Rightarrow\qquad \frac{dx}{Q(x, y)}=-\frac{dy}{P(x, y)}
\]
with homogeneous functions $P(x, y)$ and $Q(x, y)$ which satisfy to Euler's equations 
\[
x\partial_x P+y\partial_y P=\kappa P\qquad\mbox{and}\qquad
x\partial_x Q+y\partial_y Q=\kappa Q\,,\qquad \kappa\in\mathbb R\,.
\]
The corresponding partial differential equation $X(f)=0$ (\ref{pdejac-eq}) can be defined by the vector field
\begin{equation}\label{lie-eul}
X=Q(x,y)\partial_x -P(x,y)\partial_y\,,
\end{equation}
which has the following Lie bracket
\[[X,Y]=(\kappa-1)X\]
with the Euler vector field $Y=x\partial_x+y\partial_y$. In this case  
\[\Delta=\left\| 
\begin{array}{cc}
Q&-P\\
x&y\\
\end{array}
\right\|\,,\]
 multiplier 
\[
M(x,y)=\Delta^{-1}=\frac{1}{x P+yQ}\neq 0\,, 
\]
and integral 
\[
f(x,y)=
\int  \frac{P}{xP  +  yQ}dx+\int \frac{Q}{x P+yQ} dy+
\iint
\frac{(1-\kappa)QP - xP\partial_xQ-yQ\partial_y P }{( xP  +  yQ)^2}dxdy\,.
\]
were found by Euler in \S477 of his textbook \cite{eul-int}. 
Euler writes that $\varphi(f)$ could be polynomial, rational or transcendent function without explaining what this formula means. 

\section{Complete systems in $\mathbb R^3$}
\setcounter{equation}{0}
Following  Jacobi's idea we can take any two independent functions $f_{1,2}$ and compute three functional determinants  
\begin{equation*}
A_1=\left|
  \begin{smallmatrix}
    1 & 0 & 0 \\
    \partial_1f_1 & \partial_2f_1 & \partial_3 f_1 \\
    \partial_1f_2 & \partial_2f_2 & \partial_3 f_2 \\
  \end{smallmatrix}
\right|\,,\qquad
A_2=\left|
  \begin{smallmatrix}
    0 & 1 & 0 \\
    \partial_1f_1 & \partial_2f_1 & \partial_3 f_1 \\
    \partial_1f_2 & \partial_2f_2 & \partial_3 f_2 \\
  \end{smallmatrix}
\right|\,,\qquad
A_3=\left|
  \begin{smallmatrix}
    0 & 0 & 1 \\
    \partial_1f_1 & \partial_2f_1 & \partial_3 f_1 \\
    \partial_1f_2 & \partial_2f_2 & \partial_3 f_2 \\
  \end{smallmatrix}
\right|\,,
\end{equation*}
which together with the third function $M$ define vector field $X$ (\ref{jac-1}) and the corresponding complete systems of differential equations (\ref{ode-1}). The main problem here is choosing suitable functions $f_1$, $f_2$ and $M$ to construct a single-valued vector field $X$.

Following Lie's idea we can take three vector fields $Y_1,Y_2,Y_3$ 
which span three-dimensional real Lie algebra. Classification of these algebras may be found in \cite{lie-book,lie-book2} and  \cite{bia-0}.  To construct complete systems of differential equations (\ref{ode-1}) we have to find representation of the given real algebra in the space of vector fields
\[
Y_k=a_{k}(x)\partial_1+b_{k}(x)\partial_2+c_{k}(x)\partial_3\,,\qquad k=1,2,3.
\]
In his theory of function groups Lie always supposed that the functions $a_k$, $b_k$ and $c_k$ are homogeneous functions with respect to $x_1,x_2$ and $x_3$. It means that one of the vector fields $Y_1,Y_2$ or  $Y_3$ is the Euler vector field, and our aim is choosing only two suitable vector fields.  The main problem here is that the coefficients $c^k_{ij}$ are functions of the first integrals in the generic case.

\subsection{Jacobi's construction of the Darboux-Halphen system}
Let us take complete solution of the Darboux-Halphen system found by  Halphen \cite{hal,hal2} and Brioschi \cite{bri}. Solving the corresponding  equations (\ref{intjac-eq}) with respect to $n-1=2$ constants we obtain two first integrals
\begin{align}
f_1=&\frac{x_2}{\sqrt{x_2 - x_3}}\,\mathsf{K}\left(z\right)
 - \sqrt{x_2 - x_3}\,\mathsf{E}\left(z\right)\,, \qquad z=\frac{\sqrt{x_1 - x_3}}{\sqrt{x_2 - x_3}}
\nonumber\\
\label{f12-dh}\\
f_2=&\frac{x_3}{\sqrt{x_2 - x_3}}\,\mathsf{K}'\left(z\right)
 + \sqrt{x_2 - x_3}\,\mathsf{E}'\left(z\right)\,,
 \nonumber
\end{align}
Here we regard $x_1$ as independent variable in (\ref{intjac-eq}) and $\mathsf{K}(z)$ and $\mathsf{K}'(z)$ are complete and complementary complete elliptic integrals of the first kind, whereas
$\mathsf{E}(z)$ and $\mathsf{E}'(z)$ are complete and complementary complete elliptic integrals of the second kind.

\begin{prop}
Let $\mathsf{K},\mathsf{E},\mathsf{K}',\mathsf{E}'$ be complete elliptic integrals in a fixed normalization, so that
Legendre’s relation has the following form
\[\left(\mathsf{K}(z)-\mathsf{E}(z)\right)\mathsf{K}'(z) - \mathsf{K}(z)\mathsf{E}'(z)=C_L\]
In the standard convention  nonzero constant $C_L$ is equal to $-\pi/2$ \cite{as}.

After absorbing this constant into the normalization of the Jacobi multiplier,
the multiplier may be written as
\begin{equation}\label{m-dh}
M=\frac{\left(\mathsf{K}\left(z\right)-\mathsf{E}\left(z\right)\right)\mathsf{K}'(z) - \mathsf{K}(z)\mathsf{E}'(z)}{4(x_1 - x_3)(x_2 - x_3)(x_1-x_2)}=\frac{1}{4(x_1 - x_3)(x_2 - x_3)(x_1-x_2)}\,.
\end{equation}
With this normalization the Jacobi's vector field $X$ (\ref{jac-1})  associated with the first integrals $f_{1}$ and $f_2$
is equal to
\[
X=\left(x_2x_3-x_1x_2 - x_1x_3\right)\partial_1+
\left(x_1x_3-x_2x_1-x_2x_3\right)\partial_2+\left(x_1x_2-x_3x_1-x_3x_2\right)\partial_3\,.
\]
\end{prop}
This multiplier $M$ (\ref{m-dh}) coincides with the multiplier from \cite{chak03}. In the next subsection will use the generalization of the Legendre relation for hypergeometric functions \cite{kar}.

Thus, we have demonstrated that multivalued transcendental functions $f_k$ can generate a single-valued differential equation within the context of the Jacobi method.

Linear polynomials 
\begin{equation} \label{darb-1}
d_1=x_2-x_3\,,\qquad d_2=x_1-x_3\,,\qquad d_3=x_1-x_2
\end{equation}
are the Darboux polynomials of the vector field $X$ 
\[
X(d_1)=-2x_1d_1\,,\qquad X(d_2)=-2x_2d_2\,,\qquad X(d_3)=-2x_3d_3
\]
with cofactors 
\[c_1=-2x_1\,,\qquad c_2=-2x_2\,,\qquad c_3=-2x_3\,,\] respectively. Using these polynomials and cofactors we can rewrite first integrals and multiplier in the following form
\[
f_1=-\frac{c_2}{2\sqrt{d_1}}\,\mathsf{K}\left(z\right)
 - \sqrt{d_1}\,\mathsf{E}\left(z\right)\,, \qquad
f_2=-\frac{c_3}{2\sqrt{d_1}}\,\mathsf{K}'\left(z\right)
 + \sqrt{d_1}\,\mathsf{E}'\left(z\right)\,, \qquad z=\sqrt{\frac{d_2}{d_1}\,}
\]
and
\[
M=\frac{G_L}{4d_1d_2d_3}\,. 
\]
It is clear that the multiplier $M$ is also a Darboux integral
\[
X(M)=2(x_1+x_2+x_3)M
\]
with cofactor $y=2(x_1+x_2+x_3)$ which  satisfies the Chazy equation
\begin{equation}\label{chazy}
\frac{d^3 y}{d t^3} + 2y \frac{d^2 y}{d t^2} - 3\left(\frac{d y}{d t}\right)^2=0\,.
\end{equation}
This equation appears in a series of
papers by J. Chazy between in his work on the classification of third-order ordinary differential equations. Subsequently, the Chazy equation  was shown to arise in several areas of mathematical physics involving the
study of magnetic monopoles, self-dual Yang-Mills and Einstein equations, and topological
field theory, and special reductions of hydrodynamic-type equations, see details in \cite{chak99,chak03,chak18,chak25}.

\subsubsection{Possible generalizations of the Darboux-Halphen  system} 
Throughout this subsection all the parameters are assumed to be generic, so that none
of the denominators appearing in the displayed formulae vanishes; degenerate cases
require separate limiting formulae.

If we substitute  
\[
z=\frac{d_2^{\ell/2}}{d_1^{m/2}}=\frac{(x_1 - x_3)^{\ell/2}}{(x_2 - x_3)^{m/2}}\,,\qquad\mbox{where}\qquad \ell,m\in \mathbb N\,,
\]
into elliptic integrals (\ref{f12-dh}), we again obtain single valued inhomogeneous at $\ell\neq m$ polynomial differential system $X_{\ell m} (f)=0$ (\ref{jac-1})  of higher order, which can be considered as a generalization of the classical second order  Darboux-Halphen system.  For instance, if $\ell=1$ and $m=2$ we obtain the following polynomial system of differential equations
\[
\dot{x}_1=x_3^3-2x_2^3 +(2x_1 + 2x_3 + 1)x_2^2 - (x_3^2 + 3x_1 - x_3)x_2 - (2x_3^2 - 2x_1 + x_3)x_1\,,\]
and
\[
\dot{x}_2=(2x_2x_3 - x_3^2 + x_1 - x_2)(x_2 - x_3)\,,\qquad 
\dot{x}_3=(x_3^2 - x_1 + x_2)(x_2-x_3)\,.
\]
We can also replace elliptic integrals  the hypergeometric functions  ${}_2\mathsf{F}_1$ using generalizations of Legendre's relation \cite{kar}. Indeed, if we take functions
\begin{align}
f_1=A&\left(\tfrac{(x_2 - x_3)\bigl(a_2(x_1 - x_2) - a_3(x_1-x_3) + x_1\bigr)}{(x_1 - x_3)(x_1-x_2)} 
{}_2\mathsf{F}_1\left( [a,b], [c],z\right)\right.\nonumber
\\&+
\left.
\tfrac{(1 + a_1 + a_2-a_3)(1-a_1+ a_2 - a_3)}{2(1 + a_2)} {}_2\mathsf{F}_1\left( [a+1,b+1], [c+1],z\right)\right)
\nonumber\\
\label{f12-dhg}\\
f_2=A &\left(\tfrac{(x_2 - x_3)\bigl(-a_2(x_1 -x_2) - a_3(x_1 - x_3) + x_1\bigr)}{(x_1 - x_3)(x_1-x_2)} 
{}_2\mathsf{F}_1\left( [1-c + a, 1-c+b], [2-c],z\right)\right.
\nonumber\\
&+\left.
\tfrac{(1+a_1-a_2-a_3)(1-a_1-a_2-a_3)}{2(1 - a_2)} 
{}_2\mathsf{F}_1\left( [2-c+a,2-c+b], [3-c],z\right)\right)\,,
\nonumber
\end{align}
where
\[
z=\frac{x_1 - x_3}{x_2 - x_3}\,,\qquad
A=(x_2 - x_3)^{-3/2 - a_2/2 + a_3/2}\,(x_1-x_2)^{1 - a_3/2}\,(x_1 - x_3)^{1 + a_2}
\]
and
\[
a_1 = a-b\,,\qquad  a_2 = c-1\,,\qquad a_3 = c - b - a\,,
\]
we obtain the well-known generalization of the Darboux-Halphen system
\begin{equation}\label{halp-sys2}
\left\{ \begin{array}{l}
\dot{x}_1=x_2x_3-x_1(x_2+x_3)-\tau\,,\\
\dot{x}_2=x_1x_3-x_2(x_1+x_3)-\tau\,,\\
\dot{x}_3=x_1x_2-x_3(x_1+x_2)-\tau\,,
\end{array}
\right.
\end{equation}
where
\[
\tau=a_1^2(x_1 - x_3)(x_1-x_2)+a_2^2(x_2 - x_3)(x_2-x_1) + a_3^2(x_3-x_1)(x_3 - x_2)\,,
\]
see discussion in \cite{chak03}. 

In order to get another generalised Darboux-Halphen  system 
\begin{equation}\label{halp-sys3}
\left\{ \begin{array}{l}
\dot{x}_1=x_2x_3-x_1(x_2+x_3)-\alpha_1(x_1 - x_2)(x_1 - x_3)\,,\\
\dot{x}_2=x_1x_3-x_2(x_1+x_3)-\alpha_2(x_2 - x_3)(x_2 - x_1)\,,\\
\dot{x}_3=x_1x_2-x_3(x_1+x_2)-\alpha_3(x_3 - x_1)(x_3 - x_2)\,,
\end{array}
\right.
\end{equation} 
we can start with the following first integrals
\begin{align}
f_1=&B_1\Bigl(
c(x_2 - x_3)x_1 {}_2\mathsf{F}_1\left([a, b], [c], z\right) + b(x_1 - x_3)(x_1 - x_2)
 {}_2\mathsf{F}_1\left([a + 1, b + 1], [c+1],z\right)\Bigr)\,,
\label{f12-dhg2}\\
f_2=&B_2\Bigl(
(x_2 - x_3)(c-2)\left(\left(a-c+1\right)x_1 + (c-1)x_2\right)  {}_2\mathsf{F}_1\left([b-c + 1, a-c+ 1], [2 - c],z\right)\Bigr.
\nonumber\\
+&\Bigl.(x_1 - x_3)(x_1 - x_2)(b-c + 1)(a-c+ 1) {}_2\mathsf{F}_1\left([a-c+ 2, b-c+ 2], [3 - c],z\right)
\Bigr)\,,\nonumber
\end{align}
where
\[
B_1=\tfrac{(x_2 - x_3)^{-a/2 - b/2 - 1}(x_1 - x_3)^{c/2 - 1/2}(x_1 - x_2)^{a/2 + b/2 - c/2}}{b}
\]
\[
B_2=\tfrac{ (x_2 - x_3)^{c-a/2 - b/2  - 2}(x_1 - x_3)^{1/2 - c/2}(x_1 - x_2)^{a/2 + b/2 - c/2} }{(1+b-c)(1+a-c)}
\]
and
\[
\alpha_1=\frac{a-b}{a}\,,\qquad \alpha_2=\frac{c - 1}{a}\,,\qquad \alpha_3= \frac{a+b-c}{a}\,.
\]
As above, linear polynomials $d_1=x_2-x_3$, $d_2=x_1-x_3$ and $d_3=x_1-x_2$ are the Darboux polynomials of the both vector fields $X$ (\ref{halp-sys2}) and $X$ (\ref{halp-sys3})
\[
X(d_1)=-2c_1d_1\,,\qquad X(d_2)=-2c_2d_2\,,\qquad X(d_3)=-2c_3d_3
\]
with different cofactors $c_1$,$c_2$ and $c_3$ depending on parameters $\alpha_i$.

If  we substitute  
\[
z=\frac{(x_1 - x_3)^{\ell}}{(x_2 - x_3)^{m}}\,,\qquad \ell,m\in \mathbb N\,,
\]
into (\ref{f12-dhg}) or (\ref{f12-dhg2}) we obtain inhomogeneous at $\ell\neq m$ polynomial differential system $X_{\ell m} (f)=0$ (\ref{jac-1})  of higher order. We know nothing about complete solutions of these systems. 

\subsection{Lie's construction of the Darboux-Halphen system}
Let us try to generalize Lie's construction of Euler's homogeneous planar system (\ref{lie-eul}) and consider
the following vector fields 
\begin{equation}\label{efh-def}
 E=e_1(x)\partial_1+e_2(x)\partial_2+e_3(x)\partial_3\,,\qquad  H=2(x_1\partial_1+x_2\partial_2+x_3\partial_3)
\qquad
 F=a_1\partial_1+a_2\partial_2+a_3\partial_3\,,
\end{equation}
where $a_i\in\mathbb R$ and $e_j(x)$ are homogeneous functions of second degree which satisfy to the Euler equations
\begin{equation}\label{euler}
x_1\partial_1 e_j(x)+x_2\partial_2 e_j(x)+x_3\partial_3 e_j(x)=2 e_j(x)\,.
\end{equation}
Differential equations $F(f)=0$ and $H(f)$ are complete integrable and vector fields $F$ and $H$ have the necessary number of characteristic functions. Indeed, characteristic functions of the constant vector field  $ F$ are functions on linear polynomials 
\begin{equation}\label{darb-f}
d_{ij}=a_ix_j-a_jx_i\,,\qquad F(d_{ij})=0\,, F\wedge  H=\sum _{i,j=1}^3 d_{ij}\partial_i\wedge\partial_j\,,
\end{equation}
whereas characteristic functions of the vector field $H$ are homogeneous functions of zero degree. 
So, we add one unknown equation $E(f)$ to the complete differential systems.

Suppose, that vector fields (\ref{efh-def}) span real  $sl(2,\mathbb R)$ algebra, i.e. that
\begin{equation}\label{sl2}
[ H, E]= 2 E\,,\qquad [ H, F]=-2 F\,,\qquad [ E, F]=  H\,.
\end{equation} 
In this case linear polynomials $2d_{ij}$ are $2\times 2$ minors of the corresponding Lie's determinant $\Delta$ (\ref{m-lie})
\begin{equation}\label{m-gdh}
\Delta=\left\| 
\begin{array}{ccc}
e_1(x)&e_2(x)&e_3(x)\\
2x_1&2x_2&2x_3\\
a_1&a_2 &a_3\\
\end{array}
\right\|\,,
\end{equation}
in accordance with Lie's theory.

The last Lie bracket in (\ref{sl2}) gives rise to a system of partial differential equations on  functions  $e_j(x)$ 
\begin{equation}\label{lie-1}
a_1\partial_1 e_j(x)+a_2\partial_2 e_j(x)+a_3\partial_3 e_j(x)+2x_j=0\,,\qquad j=1,2,3\,.
\end{equation}
\begin{prop}
Assume  $a_1a_2a_3\neq 0$ and  $\Delta\neq 0$. The 
solutions of  (\ref{euler}-\ref{lie-1}) are
\begin{equation} \label{sl2-sol}
  e_1(x)=-\frac{x_1^2}{a_1} +g_1(x)\,,\qquad
  e_2(x)=-\frac{x_2^2}{a_2} +g_2(x)\,,\qquad
  e_3(x)=-\frac{x_3^2}{a_3}+ g_3(x)\,,
\end{equation} 
where the functions $g_j(x)$ satisfy
\[
 (x_1\partial_1+x_2\partial_2+x_3\partial_3)g_j=g_j\qquad\mbox{and}\qquad  F(g_j)=0\,,\qquad j=1,2,3\,.
\]
Thus the functions $g_j(x)$ are arbitrary smooth homogeneous functions of degree  2 depending only on the invariants 
\[d_{ik}=a_ix_k-a_kx_i\,,\]
of the constant vector field $F$.

As an example they could be homogeneous polynomials of the second order on $x_1,x_2$ and $x_3$
\[
  g_j(x)=\sum \alpha_{j}^{ik\ell m} d_{ik}d_{\ell m}\,,\qquad \alpha_{j}^{ik\ell m}\in\mathbb R\,,
\]
depending on parameters $\alpha_{j}^{ik\ell m}$.
\end{prop}
The proof consists of a direct solution of equations (\ref{euler}-\ref{lie-1}). We do not consider here other solutions when $a_1=0$ or $a_1=a_2=0$, either $\Delta=0$.

Since $\mathcal D=span\{E,F,H\}$ is the regular smooth involutive distribution we can prove the following Proposition using Jacobi-Clebsch-Deanha-Frobenius-Lie theorem \cite{cl61,cl66,dea,frob,lie-book}.
\begin{prop}
The following system of ordinary differential equations 
\begin{equation}\label{g3-eq}
\frac{\mathrm{d}x_1}{e_1(x)}=\frac{\mathrm{d}x_2}{e_2(x)}=\frac{{\mathrm d}x_3}{e_3(x)}
\end{equation}
is locally complete integrable in the sense that it has a complete solution. Here functions $e_i$ are given by (\ref{sl2-sol}) and Jacobi's multiplier is given by $M=\Delta^{-1}$ (\ref{m-gdh}).
\end{prop}

For the classical Darboux-Halphen system (\ref{darb-1})
\[
E=\sum_{i=1}^3 E_i\partial_i\,,\qquad H=2\sum_{i=1}^3x_i\partial_i\,,\qquad F=\partial_1+\partial_2+\partial_3
\]
where
\begin{align*}
E_1=x_2x_3-x_1(x_2 + x_3)\,,\quad
E_2=x_1x_3-x_2(x_1+x_3)\,,\quad
E_3=x_1x_2-x_3(x_1+x_2)\,.
\end{align*}
we have
\[
\Delta=\left\| 
\begin{array}{ccc}
E_1&E_2&E_3\\
2x_1&2x_2&2x_3\\
1&1&1\\
\end{array}
\right\|=4d_1d_2d_3\,.
\]
Here 
 \[
d_1=x_2-x_3\,,\qquad d_2=x_1-x_3\,,\qquad d_3=x_1-x_2
\]
are the Darboux polynomials (\ref{darb-1}) from the previous section. Thus, Lie's determinant multiplier
\[
M=\Delta^{-1}=\frac{1}{4d_1d_2d_3}
\]
is precisely the rational multiplier appearing in the Jacobi construction, up to constant $G_l$ associated with normalisation of the complete elliptic integrals.
 
\subsection{Characteristic functions}
According to Lie \cite{lie-book,lie-book2} we have to  consider characteristic functions of the vector fields $E,F$ and $H$ (\ref{efh-def}) and the corresponding function groups. Recall that Lie used the term "group" instead of the modern term  "algebra".

For brevity we put $a_1=a_2=a_3=1$ and consider only classical Darboux-Halphen system (\ref{halp-sys}) when
\begin{equation}\label{e-dh}
 E=\left(x_2x_3-x_1x_2 - x_1x_3\right)\partial_1+
\left(x_1x_3-x_2x_1-x_2x_3\right)\partial_2+\left(x_1x_2-x_3x_1-x_3x_2\right)\partial_3\,.
\end{equation}
A basis of characteristic functions of the Darboux-Halphen vector field $ E$ (\ref{e-dh}) is given by functions (\ref{f12-dh}) 
\begin{align*}
I_1=\frac{x_2}{\sqrt{x_2 - x_3}}\,\mathsf{K}\left(z\right)
 - \sqrt{x_2 - x_3}\,\mathsf{E}\left(z\right)\,, \qquad 
I_2=\frac{x_3}{\sqrt{x_2 - x_3}}\,\mathsf{K}'\left(z\right)
 + \sqrt{x_2 - x_3}\,\mathsf{E}'\left(z\right)\,.
 \nonumber
\end{align*}
In this case three-dimensional function group is spanned by functions $I_1,I_2$ and $\varphi_1,\varphi_2$   so that
\begin{equation}\label{wron}
 I_1\varphi_2-I_2\varphi_1=1
\end{equation}
and
\[
\begin{array}{llllll}
E(I_1)=0\,,&\,  E(I_2)=0\,,&\,  H(I_1)=I_1\,,&\,  H(I_2)=I_2\,,&
\,  F(I_1)=\varphi_1\,,&\,  F(I_2)=\varphi_2\,,\\ \\
E(\varphi_1)=I_1\,,&\, E(\varphi_2)=I_2\,,&\,  H(\varphi_1)=-\varphi_1\,,&\,  H(\varphi_2)=-\varphi_2\,,&
\,  F(\varphi_1)=0\,,&\,  F(\varphi_2)=0\,.
\end{array}
\]
Functions $\varphi_{1,2}(z)= F(I_{1,2})$ are equal to 
\[
\varphi_1(z)=\frac{\mathsf{K}(z)}{\sqrt{x_2-x_3}}\qquad\mbox{and} \qquad
\varphi_2(z)=\frac{\mathsf{K}'(z)}{\sqrt{x_2-x_3}}\,.
\]
Recall that $\mathsf{K}(z)$, $\mathsf{K}'(z)$ are complete and complementary complete elliptic integrals of the first kind. 

Characteristic functions $I_{1,2}$ of vector field  $ E$ and 
characteristic functions $\varphi_{1,2}$ of vector field $ F$ can be considered as a counterpart of standard action-angle variables since
\[
\frac{d}{dt}\varphi_1=I_1\qquad\mbox{and}\qquad \frac{d}{dt}\varphi_2=I_2\,,
\]
where derivatives by time $t$ are defined by (\ref{halp-sys}), see discussion in \cite{chak03}. Of course, instead of the Liouville torus for commutative integrable systems we have more complicated geometric object for the noncommutative integrable system.

\subsection{Possible generalizations of the Darboux-Halphen  system}
We can also obtain $sl(2,\mathbb R)$ realisation in the space of vector fields on $\mathbb R^n$. Indeed, let us  take  vector fields 
\[
E=\sum _{i=1}^n e_i(x)\partial_i\,,\qquad H=2\sum_{i=1}^n x_i\partial_i\qquad\mbox{and}\qquad
\qquad F=\sum _{i=1}^n a_i\partial_i\,,
\]
where $a_i\in\mathbb R$ and  $n$ functions $e_j(x)$  satisfy to the  Euler equations
\begin{equation}\label{euler-g}
\sum _{i=1}^n x_i\partial_i e_j(x)-2 e_j(x)=0\,,\qquad j=1,\ldots,n\,.
\end{equation}
Partial differential equations $F(f)=0$ and $H(f)$ are complete equations having $n-1$ independent characteristic functions obtained from linear polynomials $d_{ij}=a_ix_j-a_jx_i$ and rational functions $x_i/x_j$, respectively.

Suppose that vector fields $E,F$ and $H$ span real  $sl(2,\mathbb R)$ algebra (\ref{sl2}). In this case Lie bracket 
\[
[E,F]=H
\]
yields other $n$ equations on the functions  $e_j(x)$ 
\begin{equation}\label{lie-2}
\sum _{i=1}^n  a_i\partial_i e_j(x)+2x_j=0\,,\qquad j=1,\ldots,n.
\end{equation}
Equations (\ref{euler-g}-\ref{lie-2}) have the following solutions 
\[
e_k(x)=-\frac{x_k^2}{a_k}+g_k(x)\,.
\]
Here $g_k(x)$ are smooth homogeneous functions of degree two depending only on components $d_{ij}=x_ia_j-x_ja_i$ of the exterior product $F\wedge H$ so that $F(g_k)=0$. As an example  we can again consider polynomials
\[ g_k(x)=\sum \alpha_{k}^{ij\ell m} d_{ij}d_{\ell m}\,,\qquad \alpha_{k}^{ij\ell m}\in\mathbb R\,.
\]
In this case we have to compute all the minors of the $3\times 4$ matrix 
\[
L=\left(
  \begin{array}{cccc}
    e_1 & e_2 & e_3 & e_4 \\
    2x_1 & 2x_2 & 2x_3 & 2x_4 \\
    a_1 & a_2 & a_3 & a_4 \\
  \end{array}
\right)
\]
and to apply Lie's theory for the proof of completeness of the corresponding differential system.

Let us consider the simplest solution at $g_k(x)=0$
\[e_k(x)=-\frac{x_k^2}{a_k}\,.\]
In this case the multiplier of complete differential equation $E(f)=0$ can be found directly solving equation div$(ME)=0$. It is equal to
\[
M=\frac{1}{x_1^2x_2^2x_3^2x_4^2}\,,\]
so that
\[\mbox{div}(ME)=0\qquad\mbox{and}\qquad \mbox{div}(MF)\neq 0\,,\qquad
\mbox{div}(MH)\neq 0\,.\]
As a basis of characteristic functions of $E$ we can take  rational first integrals
\[
I_1=\frac{a_1x_2-a_2x_1}{x_1x_2}\,,\qquad 
I_2=\frac{a_1x_3-a_3x_1}{x_1x_3}\,,\qquad
I_3=\frac{a_1x_4-a_4x_1}{x_1x_4}\,.
\]
In the generic case $g_k\neq 0$ we know nothing about multiplier and characteristic functions.

\section{Darboux-Halphen system as the Ramanujan system}
\setcounter{equation}{0}
At the suggestion of V. Rubtsov \cite{rub}, and with his permission, we briefly revisit the direct identification of the Darboux-Halphen vector field with the Ramanujan vector field on the ring of quasimodular forms.

In \cite{ram} Ramanujan introduced the functions
\[
P(q)= 1-24\sum_{n=1}^\infty \frac{nq^n}{1-q^n}\,,\qquad
Q(q)= 1+240\sum_{n=1}^\infty\frac{n^3q^n}{1-q^n}\,,\qquad
R(q)= 1-504\sum_{n=1}^\infty\frac{n^5q^n}{1-q^n}
 \]
and showed that these functions satisfy the following equation
\begin{equation}\label{ram-1}
q\frac{\mathrm{d}P}{\mathrm{d}q}=\frac{P^{2}-Q}{12}, \qquad q\frac{\mathrm{d}Q}{\mathrm{d}q}=\frac{P Q-R}{3}, \qquad q\frac{\mathrm{d}R}{\mathrm{d}q}=\frac{P R-Q^{2}}{2} \,.
\end{equation}
In modern terminology the functions $P, Q$ and $R$ are the well-known Eisenstein series
$E_2, E_4$ and $E_6$ for the modular group $PSL_2(\mathbb Z)$, see  \cite{chak25,taht92,rub26} and references within.

Following to \cite{rub}, we introduce the elementary symmetric functions of the $x_1,x_2$ and $x_3$ variables:
\[
\sigma=x_{1}+x_{2}+x_{3}, \quad \pi=x_{1} x_{2}+x_{1} x_{3}+x_{2} x_{3}, \quad \rho=x_{1} x_{2} x_{3}
\]
A direct computation from (\ref{halp-sys}) yields
\begin{equation}\label{v1}
\dot{\sigma}=-\pi, \quad \dot{\pi}=-6 \rho, \quad \dot{\rho}=\pi^{2}-4 \sigma \rho 
\end{equation}
Now define new variables by the polynomial map, with a free nonzero scale parameter $\alpha$ 
\begin{equation}\label{v2}
P=\alpha \sigma, \quad Q=\alpha^{2}\left(\sigma^{2}-3 \pi\right), \quad R=\alpha^{3}\left(\sigma^{3}-\frac{9}{2} \sigma \pi+\frac{27}{2} \rho\right) 
\end{equation}
and the rescaled derivation by time  
\begin{equation}\label{time-rep}
\mathcal{D}=-\frac{\alpha}{4} \frac{\mathrm{~d}}{\mathrm{~d} t}.
\end{equation}
 One verifies by direct substitution into (\ref{v1}) that $P$, $Q, R$ satisfy exactly the Ramanujan system (\ref{ram-1})
\begin{equation*}
\mathcal{D}(P)=\frac{P^{2}-Q}{12}, \quad \mathcal{D}(Q)=\frac{P Q-R}{3}, \quad \mathcal{D}(R)=\frac{P R-Q^{2}}{2} \,.
\end{equation*}
This is an exact algebraic equivalence: no approximation or passage to a limit is involved. The proof is a term-by-term verification; we record the three checks for completeness. 
\par\noindent
First equation:
\[
\mathcal{D}(P)=\alpha\left(-\frac{\alpha}{4}\right) \dot{\sigma}=\frac{\alpha^{2} \pi}{4}, \quad \frac{P^{2}-Q}{12}=\frac{\alpha^{2} \sigma^{2}-\alpha^{2}\left(\sigma^{2}-3 \pi\right)}{12}=\frac{\alpha^{2} \pi}{4}\,.
\]
\par\noindent
Second equation:
\[
\mathcal{D}(Q)=-\frac{\alpha^{3}}{4}(2 \sigma \dot{\sigma}-3 \dot{\pi})=-\frac{\alpha^{3}}{4}(-2 \sigma \pi+18 \rho)=\frac{\alpha^{3}(\sigma \pi-9 \rho)}{2}\,,
\]
and
\[
\frac{P Q-R}{3}=\frac{\alpha^{3}\left[\sigma\left(\sigma^{2}-3 \pi\right)-\sigma^{3}+\frac{9}{2} \sigma \pi-\frac{27}{2} \rho\right]}{3}=\frac{\alpha^{3}(\sigma \pi-9 \rho)}{2}\,.
\]
\par\noindent
Third equation:
\[
\begin{gathered}
\mathcal{D}(R)=-\frac{\alpha^{4}}{4}\left(3 \sigma^{2} \dot{\sigma}-\frac{9}{2} \sigma \dot{\pi}+\frac{27}{2} \dot{\rho}\right)=\frac{3 \alpha^{4}}{4}\left(\sigma^{2} \pi-6 \pi^{2}+9 \sigma \rho\right) \\
\frac{P R-Q^{2}}{2}=\frac{\alpha^{4}\left(\frac{3}{2} \sigma^{2} \pi+\frac{27}{2} \sigma \rho-9 \pi^{2}\right)}{2}=\frac{3 \alpha^{4}}{4}\left(\sigma^{2} \pi-6 \pi^{2}+9 \sigma \rho\right)
\end{gathered}
\]
The inverse map reads as
\begin{equation*}
\sigma=\frac{P}{\alpha}, \qquad \pi=\frac{P^{2}-Q}{3 \alpha^{2}}, \qquad \rho=\frac{2 R+P^{3}-3 P Q}{27 \alpha^{3}} \,.
\end{equation*}
The original variables $x_{1}, x_{2}, x_{3}$ are recovered as the three roots of the depressed cubic, i.e. the cubic with no  $W^{2}$-term
\begin{equation*}
W^{3}-\frac{Q}{3} W-\frac{2 R}{27}=0, \quad W_{i}=\alpha\left(x_{i}-\frac{\sigma}{3}\right) 
\end{equation*}
whose discriminant $4\left(Q^{3}-R^{2}\right) / 27$ equals $\alpha^{6}\left(d_{1} d_{2} d_{3}\right)^{2}$ in terms of the Darboux polynomials (\ref{darb-1}) or characteristic function (\ref{darb-f}) of the Lie vector field $F$. Consequently,
\begin{equation} \label{v6}
Q^{3}-R^{2}=\frac{27 \alpha^{6}}{4}\left(d_{1} d_{2} d_{3}\right)^{2} 
\end{equation}
which, together with the Jacobi-Lie multiplier 
\[M=\frac{1}{4 d_{1} d_{2} d_{3}}\,,\] 
yields the explicit dictionary
\begin{equation}\label{v7}
M=\frac{\sqrt{27} \alpha^{3}}{8 \sqrt{Q^{3}-R^{2}}} \,.
\end{equation}
In the standard normalisation $Q=E_{4}, R=E_{6}$ we have \[Q^{3}-R^{2}=1728\, \Delta(\tau),\] where $\Delta$ is the modular discriminant, so we have formally  \[M \propto \Delta(\tau)^{-1 / 2}.\]
So, the Jacobi-Lie multiplier $M$ transforms formally as an object of modular weight
weight -6, after choosing a branch $\Delta(\tau)^{1/2}$ and fixing normalization.

The preceding transformation is purely algebraic and contains an arbitrary nonzero
arbitrary parameter $\alpha$. To compare it with the standard modular derivation 
\[\mathcal D=\frac{1}{2\pi i}\frac{\mathrm{d}}{\mathrm{d}\tau}=q\frac{\mathrm{d}}{\mathrm{d} q}\]
 we must fix the scale using the elliptic period normalization. The Wronskian
condition (\ref{wron})
\[I_{1} \varphi_{2}-I_{2} \varphi_{1}=1\]
implies
\[
\frac{\mathrm{d} \tau}{\mathrm{~d} t}=-\frac{x_{2}-x_{3}}{K^{2}(z)}=-\frac{d_{1}}{K^{2}(z)}
\]
Consequently, in the standard modular time we have
\[\mathcal{D}=-\frac{\alpha}{4}\frac{ \mathrm{d}}{\mathrm{d} t}\,,\qquad \alpha=\frac{2 K^{2}}{\pi i d_{1}}\,,\] 
up to chosen normalization of the complete elliptic integrals.

Summing up, the Ramanujan derivation $q \mathrm{~d} / \mathrm{d} q$ is precisely the Darboux-Halphen vector field written in the symmetric coordinates and expressed in the modular time $\tau$.

\textbf{Remark 1.} The Chazy equation for $y=2 \sigma$ (\ref{chazy}) is an immediate corollary. In the Ramanujan normalisation, $P=E_{2}$ satisfies the Chazy equation as a function of $\tau$ (a classical fact), and the relation $P=\alpha \sigma$ with $\mathcal{D} \propto \mathrm{d} / \mathrm{d} t$ transfers this to the Darboux-Halphen time $t$ with no additional work. The Chazy equation is thus the unique scalar invariant shared by the Darboux-Halphen vector field and the Ramanujan vector field, and the identification (\ref{v2}) is the precise algebraic reason for this coincidence.

\textbf{Remark 2.} Identity (\ref{v7}) also provides a natural explanation for the well-known relation between the Darboux polynomials and the modular discriminant, and clarifies the sense in which the Darboux-Halphen system is modularly integrable. The Jacobi-Lie multiplier $M$ transforms formally as an object of modular weight
weight -6 (up to the factor $\alpha^{3}$ ), while the solutions  $x_{i}$ are weight -2 quasimodular forms.

The formula for $Q$ has a further pleasant form in terms of the Darboux polynomials (characteristic functions). Since $\left(x_{1}-x_{2}\right)^{2}+\left(x_{1}-x_{3}\right)^{2}+\left(x_{2}-x_{3}\right)^{2}=2\left(\sigma^{2}-3 \pi\right)$, one has
\begin{equation*}
Q=\frac{\alpha^{2}}{2}\left(d_{1}^{2}+d_{2}^{2}+d_{3}^{2}\right) 
\end{equation*}
so the Eisenstein series $E_{4}$ is (up to scale) the sum of squares of the Darboux polynomials, while $E_{6}$ is determined by their product through (\ref{v6}).

\section{Conclusion}
Fundamental contributions of Jacobi and Lie form the basis of classical  theory of complete integrable, incomplete integrable and nonintegrable systems of differential equations \cite{for90}. Of course, we have to remember the keystones of this theory to study a rich behavior of  physically important differential equations.  

In this note we recover Darboux-Halphen system of equations in the framework of the Jacobi and Lie theories of complete Pfaffian equations. Our main aim is to prove that both these theories can be useful to construct various generalizations of differential systems with known complete solutions.

After discussing the simple real algebra $sl(2, \mathbb R)$, it makes sense to move on to the algebra $sl(3)$ within the framework of Lie theory and to consider solvable (integrable in Lie's notation) algebras \cite{ts26} when a part of the first integrals are globally defined functions whereas other integrals are multi-valued functions.

In \cite{chak25}, the so-called Darboux-Halphen system of order 9 (DH9), its fifth-order reduction system (DH5), and the Chazy and Ramamani equations are discussed. It would be also interesting to describe these DH9 and DH5 systems within the framework of a Jacobi or Lie construction, analogous to those in this note.

\vskip0.2truecm
The research  was carried out with the financial support of the Ministry of Science and Higher Education of the Russian Federation in the framework of a scientific project under agreement N 075-15-2025-013 by St.Petersburg State University as part of the national project “Science and Universities” in 2025.

\vskip0.2truecm
\textbf{Conflict of Interests:}  The author declares that he has no conflict of interest.

\end{document}